\documentstyle[12pt]{article}

\begin{document}
\baselineskip=24pt
\pagestyle{plain}
\title{
\begin{flushright}
CCNY-HEP-92/7  \\
\end{flushright}
Various condensed matter Hamiltonians\\ in terms of
U(2/2) operators\\ and their symmetry structures}
\author{Ko Okumura\footnote{On leave from
 Department of Physics, Faculty of Science and Technology,
Keio University, Yokohama 223, Japan}\\
{\em Department of Physics, City College of}\\
{\em the City University of New York, New York, NY10031}}

\maketitle
\begin{abstract}
We rewrite various lattice Hamiltonian in condensed matter physics
in terms of U(2/2) operators that we introduce. In this
representation the symmetry structure of the models becomes clear.
Especially, the Heisenberg, the supersymmetric t-J and a newly proposed
high-$T_c$ superconducting Hamiltonian reduce to the same form
$H=$$-t\sum_{<jk>}\sum_{ac}X^{ac}_j X^{ca}_k (-1)^{F(c)}$.
This representation also gives us a systematic way of searching for
the symmetries of the system.
\end{abstract}
\newpage

\section{INTRODUCTION}
\label{s1}

Recently the study of strongly correlated electrons on a lattice has become
one of the central problems in solid state physics in connection with
high-$T_c$ superconductivity.
In this paper we concentrate on the symmetry structures of those Hamiltonians
which have been proposed as models of
high-$T_c$ superconductivity, that is, the Hubbard, Heisenberg, t-J models
and a new
model recently proposed by F.\ Essler,
V.\ Korepin and K.\ Schoutens (EKS model)\cite{EKS}.

The innovation made here is the introduction of $\gamma$ (Clifford) operators.
The lattice Hamiltonians are usually described by $c^{\pm}_{j\sigma}$
which form a basis of the fundamental representation of U(2).
It is known that 4 real and imaginary components of $c_{j\uparrow}$ and
$c_{j\downarrow}$ (see (\ref{2-1}))
form a 4-dimensional Clifford algebra $\gamma^{\mu}$\cite{MS}.
Through multiplications of the $\gamma$'s one can construct a maximal set
of 16 independent operators which form U(2/2) algebra\cite{KO,SU}.
Then it is  natural that the U(2/2) symmetry of the system should
manifest itself
if we express the model in terms of these U(2/2) operators.

The U(2/2) operators are explicitly given in terms of the $t$- and $X$-
operators in Sec.~II.
%We re-express these models in terms of $\gamma$- and  $X$-operators in
%Sec.~II\cite{MS}.
Both of them are the generators of the U(2/2) superalgebra
in different bases which is essentially {\em generated} by the 4-dimensional
Clifford algebra.%Taking this into account, we call $\gamma$- and
%$X$-operators U(2/2)=4D Clifford operators collectively.

With the U(2/2) operator representation some of the symmetry
structures of the models become transparent. This representation also
gives a systematic method of finding out the symmetries of the system.

Supergroup U(M/N) is quite important in condensed matter physics.
It is a unitary transformation of the states assigned to each site.
Indeed the Heisenberg model has U(2) symmetry, t-J model U(1/2) and
the new model U(2/2). Notice here that these models with U(M/N)
symmetry are all solvable at least in one-dimension. If we want a
solvable {\em multi-band} model we may need a higher
U(M/N)\cite{KO}.

Now let us specify the models we deal with explicitly.
The first one is the Hubbard model.
\begin{equation}
\label{Hubb}
H_{Hubb.}=-2t\sum_{<jk>}\sum_{\sigma=\uparrow,\downarrow}
c^{+}_{j\sigma} c_{k\sigma}+U\sum_{j=1}^L n_{j\uparrow} n_{j\downarrow}
-\mu\sum_{j=1}^L n_j
\end{equation}
where operator
$c^{\pm}_{j\sigma}(c^{-}_{j\sigma}\equiv c_{j\sigma};j=1 \cdots L)$
defined on each site (L : total number of lattice site) satisfy
anti-commutation relation
\begin{equation}
\label{acom}
\{c_{j\sigma},c^{+}_{j'\sigma'}\}=\delta_{jj'} \delta_{\sigma\sigma'}
\quad ; \quad \mbox{others}=0,
\end{equation}

The number operator $n_j$ is defined by
$n_j=\sum_\sigma n_{j\sigma}$
($n_{j\sigma}=c^{+}_{j\sigma} c_{j\sigma}$)
and $\mu$ is the chemical potential.

In what follows $\sum_{<jk>}$ implies the summation over the nearest
neighbors in which the pairs $(j,k)$ and ($k,j$) have to be counted once
each so that the summand can be always considered as symmetric under
interchanging $j\leftrightarrow k$.

If we set the chemical potential $\mu$ to be $U/2$, the system has the
SO(4) symmetry\cite{Y,YZ}. This condition is satisfied at
half-filling\cite{Pelizz}. One of the subgroups of  SO(4) is the spin
SU(2) formed by the spin operators
$S^{\pm}=\sum_{j=1}^L S^{\pm}_j$,
$S^{z}=\sum_{j=1}^L S^{z}_j$ such that
\begin{equation}
\label{Sspin}
S^{-}_j =c^{+}_{j\uparrow}c_{j\downarrow}
, \quad
S^{+}_j =c^{+}_{j\downarrow}c_{j\uparrow}
, \quad
S^{z}_j = \frac{1}{2}(n_{j\uparrow}-n_{j\downarrow})
; \quad
S^{\pm}_j =S^{x}_j \pm i S^{y}_j
\end{equation}
Another subgroup of this SO(4) symmetry is the $\eta$ pairing symmetry
recently pointed out by C.\ N.\ Yang and S.\ C.\ Zhang\cite{YZ}.

Next we consider the Heisenberg model defined by
\begin{equation}
\label{Heisen}
H_{Heisen.}^S=-J\sum_{<jk>}(\vec{S}_j\cdot\vec{S}_k-\frac{1}{4})
\end{equation}
and the t-J model
\begin{equation}
\label{t-J}
H^{eff}_{t-J}=-2t\sum_{<jk>}\sum_{\sigma}c^{+}_{j\sigma} c_{k\sigma}
+J\sum_{<jk>}(\vec{S}_j\cdot\vec{S}_k-\frac{n_j n_k}{4})
-\mu\sum_{j=1}^L n_j
\end{equation}
where we have used the notation
$\vec{S}_j\cdot\vec{S}_k =\sum_{\alpha=x,y,z}S^{\alpha}_j S^{\alpha}_k$.
Notice that $H^{eff}_{t-J}$ acts on the Hilbert space where no double
occupancy of sites are allowed. In other words  the true t-J Hamiltonian is
$H_{t-J}=PH^{eff}_{t-J}P$ where $P$ is the Gutzwiller projection operator;
$P=\prod_{j=1}^LP_j=\prod_{j=1}^L(1-n_{j\uparrow}n_{j\downarrow})$.
Setting
 $\mu/(2z)=2t=J$
with $z$ the number of the nearest neighbor sites of each site, we get
the U(1/2) supersymmetric t-J model\cite{t-J}.

There are close relations among the above three models,
that is,
$H_{Hubb.}$ reduces to $H_{Heisen.}$ in the strong coupling limit
$(U/t>>1)$ at half-filling\cite{Hubb} and also reduces to
$H_{t-J}$ in the large $U$ limit in a certain sense\cite{Hirsh}.
But the supersymmetric t-J model (2t=J) itself can not be obtained
from $H_{Hubb.}$

The final example is the EKS model\cite{EKS}.
This model  describes strongly correlated electrons on a general
d-dimensional lattice as a new exactly solvable model for high-$T_c$
superconductivity. This model has many interesting features but here
we focus our attention on its symmetry structure U(2/2).
One of the subgroup is SU($2)_\eta$ formed by $\eta$-spin operators;
\begin{equation}
\label{etaspin}
\eta^{-}_j =c_{j\uparrow}c_{j\downarrow}
, \quad
\eta^{+}_j =c^{+}_{j\downarrow}c^{+}_{j\uparrow}
, \quad
\eta^{z}_j = -\frac{1}{2}(n_{j}-1)
; \quad
\eta^{\pm}_j  =\eta^{x}_j \pm i \eta^{y}_j.
\end{equation}
Then (the supersymmetric part of) the EKS Hamiltonian is given by
\begin{equation}
\label{EKS}
H_{EKS}=-2t\sum_{<jk>}\{\sum_{\sigma}
c^{+}_{j\sigma}c_{k\sigma}(1-n_{j,-\sigma}-n_{k,-\sigma})
+\vec{\eta}_j\cdot\vec{\eta}_k-\vec{S}_j\cdot\vec{S}_k
+(n_{j\uparrow}-\frac{1}{2})(n_{j\downarrow}-\frac{1}{2})\}.
\end{equation}

In the above definitions of the Hamiltonians we have included the chemical
potential term only if such a term contributes to bringing the system
higher symmetries.

\setcounter{equation}{0}
\section{VARIOUS HAMILTONIANS IN TERMS OF U(2/2) OPERATORS[2]}
\label{s2}

At each site we introduce the $\gamma$-operators
\begin{eqnarray}
\label{2-1}
\gamma^1_j=c_{j\uparrow}+c_{j\uparrow}^+
& &
\gamma^2_j=-i(c_{j\uparrow}-c_{j\uparrow}^+)
\nonumber
\\
\gamma^3_j=c_{j\downarrow}+c_{j\downarrow}^+
& &
\gamma^4_j=-i(c_{j\downarrow}-c_{j\downarrow}^+)
\end{eqnarray}
which satisfy $\{\gamma^{\mu}_j,\gamma^{\nu}_k\}=2\delta_{\mu\nu}\delta_{jk}$.
We further introduce $\gamma^5_j=-\gamma^1_j\gamma^2_j\gamma^3_j\gamma^4_j$
and $\sigma^{\mu\nu}_j=[\gamma^{\mu}_j,\gamma^{\nu}_j]/(2i)$.
Then (as we see explicitly in (\ref{2-7})) the 16 operators
$t^{\alpha}_j=(1_j,\gamma^{\mu}_j,\sigma^{\mu\nu}_j,
i\gamma^5_j\gamma^{\mu}_j, \gamma^5_j)$
form the superalgebra U(2/2) where $\gamma^{\mu}_j$ and
$i\gamma^5_j\gamma^{\mu}_j$ are the  Fermi operators while
$1_j$, $\sigma^{\mu\nu}_j$, and $\gamma^5_j$ Bosonic operators.
Notice here that the local $t^{\alpha}_j$ can also form U(4),
U(3/1) and U(1/3) but the
global $t^{\alpha}=\sum_{j=1}^L t^{\alpha}_j$ forms only U(2/2).
With these operators the Hamiltonian becomes\cite{R1} (omitting constant terms)
\begin{equation}
\label{2-2}
H_{Hubb.}=-i\sum_{<jk>}(\gamma^1_j\gamma^2_k+\gamma^3_j\gamma^4_k)
+\frac{U}{4}\sum_j\gamma^5_j+(\frac{\mu}{2}-\frac{U}{4})
\sum_j(\sigma^{12}_j+\sigma^{34}_j),
\end{equation}
\begin{equation}
\label{2-3}
H^S_{Heisen.}=-\frac{J}{2^5}\sum_{<jk>}(1-\gamma^5_j)
\sigma^{\mu\nu}_j\sigma^{\mu\nu}_k,
\end{equation}
\begin{equation}
\label{2-4}
H_{EKS}=\frac{t}{2}\sum_{<jk>}\gamma^5_j(\gamma^{\mu}_j\gamma^{\nu}_k
-\frac{1}{4}\sigma^{\mu\nu}_j\sigma^{\mu\nu}_k-1).
\end{equation}
In the above expression, noting that $\mu$ and $\nu$ corresponds to O(4)
vector or tensor indices, some O(4) invariances of the models are transparent
as we specify in Sec.~III, that is, if there are no un-contracted indices
$\mu$ or $\nu$, the term is O(4) invariant.

Next we consider the following 16 hermitian  matrices
$T^{\alpha}\equiv(1, \Gamma^{\mu}, \Sigma^{\mu\nu}, iA^{\mu}, \Gamma^5)$
$(\Sigma^{\mu\nu}=-\Sigma^{\nu\mu})$ which correspond to a $4\times 4$ matrix
representation of $t^{\alpha}_j$ or matrix of the fundamental (defining)
representation of U(2/2) with
$tr(T^{\alpha}T^{\beta})=4\delta_{\alpha\beta}$.
\begin{eqnarray}
\Gamma^1=\left(\matrix{O&I\cr I&O\cr}\right)
\quad
\Gamma^2=\left(\matrix{O&-i\sigma_3\cr
                       i\sigma_3&O\cr}\right)
&\quad&
\Gamma^3=\left(\matrix{O&i\sigma_2\cr
                       -i\sigma_2&O\cr}\right)
\quad
\Gamma^4=\left(\matrix{O&-i\sigma_1\cr
                       i\sigma_1&O\cr}\right)
\nonumber
\\
iA^1=\left(\matrix{O&iI\cr -iI&O\cr}\right)
\quad
iA^2=\left(\matrix{O&\sigma_3\cr
                       \sigma_3&O\cr}\right)
&\quad&
iA^3=\left(\matrix{O&-\sigma_2\cr
                       -\sigma_2&O\cr}\right)
\quad
iA^4=\left(\matrix{O&\sigma_1\cr
                       \sigma_1&O\cr}\right)
\nonumber
\\
\Sigma^{12}=\left(\matrix{\sigma_3&O\cr
                       O&-\sigma_3\cr}\right)
\quad
\Sigma^{13}=\left(\matrix{-\sigma_2&O\cr
                       O&\sigma_2\cr}\right)
&\quad&
\Sigma^{14}=\left(\matrix{\sigma_1&O\cr
                       O&-\sigma_1\cr}\right)
\nonumber
\\
\Sigma^{34}=\left(\matrix{\sigma_3&O\cr
                       O&\sigma_3\cr}\right)
\quad
\Sigma^{24}=\left(\matrix{\sigma_2&O\cr
                       O&\sigma_2\cr}\right)
&\quad&
\Sigma^{23}=\left(\matrix{\sigma_1&O\cr
                       O&\sigma_1\cr}\right)
\nonumber
\\
\Gamma^5=\left(\matrix{I&O\cr
                       O&-I\cr}\right)
\quad
1=\left(\matrix{I&O\cr
                       O&I\cr}\right)
\end{eqnarray}
Here $\sigma_i$'s are $2\times 2$ Pauli-matrices, $I$ and $O$ are
 $2\times 2$ unit and zero matrices respectively.
 Note here that $\Gamma^{\mu}$ and $iA^{\mu}$ take the form of the
Fermi block matrix while the others Bose block matrix (see P. Freund
in \cite{SU}).
Again, notice that $T^{\alpha}$
can form U(4) etc.\ other than U(2/2) though the glogal $t^{\alpha}$ form only
U(2/2).
Now the $X^{ac}_j$ operators are defined as  another different basis of the
same U(2/2) algebra, or a {\em linear combination} of
the $t^{\alpha}_j$'s;
\begin{equation}
\label{2-6}
X^{ac}_j=\frac{1}{4}T^{\alpha}_{ca}t^{\alpha}_j \quad \mbox{or} \quad
t^{\alpha}_j=T^{\alpha}_{ac}X^{ac}_j
\end{equation}
where $T^{\alpha}_{ca}$ is the $(c, a)$ element of the $4\times 4$
matrix and Einstein's sum rule (as we will use in what follows implicitly)
is implied.
Notice that $a,c=1,2$ correspond to the Bose sector while $a,c=3,4$
to the Fermi sector.
 As clear from the definition, $X^{ac}_j$ forms
an U(2/2) algebra as $t^{\alpha}_j$ does, which is very similar to
the U(N) algebra;
\begin{equation}
\label{2-7}
[X^{ac}_j,X^{a'c'}_{j'}]_{\pm}=\delta_{jj'}(X^{ac'}_j\delta_{a'c}
\pm X^{a'c}_j\delta_{ac'})
\end{equation}
where $[$ , $]_\pm$ represents commutator($-$) or anti-commutator(+) the
latter occurring only if both operators are fermionic\cite{R2}.
%Thus as mentioned before $\gamma$ and $X$ are essentially the same thing;
%U(2/2)=4D Clifford operator. (\ref{2-6}) is the transformation between
%the two basis.

$X^{ac}_j$ is essentially the Hubbard projection operator\cite{KO}.
Especially {\em at each site} we can set
$(X^{ac}_j)_{mn}=\delta_{ma}\delta_{nc}$ which is helpful in the
actual calculation.

With the $X$ operators, EKS, supersymmetric t-J\cite{R4}, and Heisenberg
$(J=-2t)$
model become
\begin{equation}
\label{2-8}
H_{EKS}=-t\sum_{<jk>}\sum_{ac=1,2,3,4}X^{ac}_j X^{ca}_k (-1)^{F(c)},
\end{equation}
\begin{equation}
\label{2-9}
P\tilde{H}_{t-J}^{eff}P=-t\sum_{<jk>}\sum_{ac=1,3,4}
X^{ac}_j X^{ca}_k (-1)^{F(c)}=PH_{EKS}P,
\end{equation}
\begin{equation}
\label{2-10}
P_F H_{Heisen.}^S P_F=-t\sum_{<jk>}\sum_{ac=3,4}
X^{ac}_j X^{ca}_k (-1)^{F(c)}
=P_F H_{EKS} P_F,
\end{equation}
\begin{equation}
\label{2-11}
P_B H_{Heisen.}^{\eta} P_B=-t\sum_{<jk>}\sum_{ac=1,2}
X^{ac}_j X^{ca}_k (-1)^{F(c)}
=P_B H_{EKS} P_B,
\end{equation}
where $F(b)=0$ ($F(f)=1$) with $b=1,2$ ($f=3,4$) and $H_{Heisen.}^{\eta}$
is the Heisenberg model written not in terms of the $S$-spin operator
but in terms of the
$\eta$-spin operator. We have introduced two projection operators
$P_F$ and $P_B$, $P_F$ for the two singly occupied states
$c^{+}_{j\uparrow}|0\rangle$, $c^{+}_{j\downarrow}|0\rangle$ and
$P_B$ for the
non- or doubly-occupied states
$|0\rangle$, $c^{+}_{j\uparrow}c^{+}_{j\downarrow}|0\rangle$;
\begin{equation}
\label{2-12}
P_F=\prod_{j=1}^{L}P_F^j\>; \quad P_F^j=1-(n_{j\uparrow}-1)(n_{j\downarrow}-1)
-n_{j\uparrow}n_{j\downarrow}=(1-\gamma^5_j)/2
\end{equation}
\begin{equation}
\label{2-13}
P_B=\prod_{j=1}^{L}P_B^j\>; \quad P_B^j=1+(n_{j\uparrow}-1)n_{j\downarrow}
+(n_{j\downarrow}-1)n_{j\uparrow}=(1+\gamma^5_j)/2
\end{equation}
In the above expression, we can see clearly the fact that the supersymmetric
t-J model and Heisenberg models correspond to certain subsectors of
the EKS model
as pointed out in \cite{EKS}.
In addition, noting that various subelements of $X^{ac}$
form U(M/N), several U(M/N) invariances of the models manifest themselves
as we discuss more extensively in Sec.~III.

\setcounter{equation}{0}
\section{A SYSTEMATIC ANALYSIS OF THE SYMMETRIES OF THE MODELS}
\label{s3}

If we want to search for the symmetry of the system in question
described by a one-band lattice Hamiltonian as in our models, a general
form of symmetry operators can be written as
\begin{equation}
\label{3-1}
{\cal S}=\sum_{l=1}^{L}(V^{\rho}_{l}\gamma^{\rho}_{l}+
T^{\rho\tau}_{l}\sigma^{\rho\tau}_{l}
+iA^{\rho}_{l}\gamma^{5}_{l}\gamma^{\rho}_{l}+P_l\gamma^{5}_{l})
\end{equation}
or
\begin{equation}
\label{3-2}
\tilde{{\cal S}}=\sum_{l}x^{ac}_{l}X^{ac}_{l}
\end{equation}
where $V, T, A, P$ or $x$ are complex numbers with
$T^{\rho\tau}_{l}=-T^{\tau\rho}_{l}$.

Calculating the commutaitor between $\cal S$ or $\tilde{\cal S}$
and the Hamiltonian in question, we can find the conditions which
have to be satisfied by the coefficients $V, T, A, P$ or $x$ to
render the operator  $\cal S$ or $\tilde{\cal S}$ to be the symmetry
of the system.

First examine the Hubbard model at half-filling $(\mu=U/2)$;
\begin{eqnarray}
\label{3-3}
[ {\cal S} , H_{Hubb.} ]  = & -2it\sum_{<jk>}\{ &
(V^{1}_{j}\gamma^{2}_{k}-V^{2}_{j}\gamma^{1}_{k}
+V^{3}_{j}\gamma^{4}_{k}-V^{4}_{j}\gamma^{3}_{k})
\nonumber
\\
& & -2(T^{12}_{j}-T^{12}_{k})
(\gamma^{1}_{j}\gamma^{1}_{k}+\gamma^{2}_{j}\gamma^{2}_{k})
-2(T^{34}_{j}-T^{34}_{k})
(\gamma^{3}_{j}\gamma^{3}_{k}+\gamma^{4}_{j}\gamma^{4}_{k})
\nonumber
\\
& & -4(T^{14}_{j}+T^{23}_{k})
(\gamma^{4}_{j}\gamma^{2}_{k}+\gamma^{1}_{j}\gamma^{3}_{k})
-4(T^{13}_{j}-T^{24}_{k})
(\gamma^{3}_{j}\gamma^{2}_{k}-\gamma^{1}_{j}\gamma^{4}_{k})
\nonumber
\\
& & +A^{\mu}_{j}\gamma^{5}_{j}
(\sigma^{1\mu}_{j}\gamma^{2}_{k}+\sigma^{\mu 2}_{j}\gamma^{1}_{k}
+\sigma^{3\mu}_{j}\gamma^{4}_{k}+\sigma^{\mu 4}_{j}\gamma^{3}_{k})
\nonumber
\\
& & +P_j\gamma^{5}_{j}
(\gamma^{1}_{j}\gamma^{2}_{k}-\gamma^{2}_{j}\gamma^{1}_{k}
+\gamma^{3}_{j}\gamma^{4}_{k}-\gamma^{4}_{j}\gamma^{3}_{k})\}
\nonumber
\\
& -\frac{U}{2}\sum_j &
(V^{\mu}_{j}\gamma^{5}_{j}\gamma^{\mu}_{j}+A^{\mu}_{j}\gamma^{\mu}_{j})
\end{eqnarray}

At first sight, in the strong coupling limit ($t=0$), the system has a
local O(4)$\sim$SU(2$)_S\otimes$SU(2$)_{\eta}$ whose generators are
$\sigma^{\mu\nu}_{l}\sim(\vec{S}_{l}, \vec{\eta}_{l})$
and the Hamiltonian of the system commutes with $\gamma^{5}_{l}$.
This implies the local U(2$)_S\otimes$U(2$)_{\eta}$
symmetry whose generators are
$((\vec{S}_{l}, (1-\gamma^5_{l})/2), (\vec{\eta}_{l}, (1+\gamma^5_{l})/2))$.
 For general $U$ and $t$, we notice that, by setting
$T^{12}_{j}=T^{12}_{k}=$const and the other coefficient to be zero,
$\sigma^{12}=\sum_{j=1}^{L}\sigma^{12}_{j}$ commutes with $H_{Hubb.}$
$(\mu=U/2)$. In a similar way we can find other symmetry operators.
Let us summarize in the following way.
\begin{enumerate}
\item $T^{12}_{j}=T^{12}_{k}=$ const, others = 0
\quad$\Rightarrow$\quad  $\sigma^{12}$
\item $T^{34}_{j}=T^{34}_{k}=$ const, others = 0
\quad$\Rightarrow$\quad  $\sigma^{34}$
\item $T^{14}_{j}=-T^{23}_{k}=$ const, others = 0
\quad$\Rightarrow$\quad
 $\sigma^{14}-\sigma^{23}$
\item $T^{14}_{j}=T^{23}_{j}=e^{i\vec{\pi} \cdot \vec{j}}$, others = 0
\quad$\Rightarrow$\quad
$\eta^{x}_{Y}\equiv \sum_j
e^{i\vec{\pi} \cdot \vec{j}}(\sigma^{14}_{j}+\sigma^{23}_{j})$
\item $T^{13}_{j}=T^{24}_{k}=$ const, others = 0
\quad$\Rightarrow$\quad
$\sigma^{13}+\sigma^{24}$
\item $T^{13}_{j}=-T^{24}_{j}=e^{i\vec{\pi} \cdot \vec{j}}$, others = 0
\quad$\Rightarrow$\quad
$\eta^{y}_{Y}\equiv \sum_j
e^{i\vec{\pi} \cdot \vec{j}}(\sigma^{13}_{j}-\sigma^{24}_{j})$
\end{enumerate}
 4 and 6 may need some explanation. Here we have assumed the lattice constant
to be 1 merely for simplicity. $\vec{\pi}$ denotes the d-dimensional vector
whose elements are all $\pi$ and $\vec{j}$ is the d-dimensional integer vector
which implies one site on the lattice. Then for the nearest neighbors $j, k$
the relation
$T^{14}_{j}+T^{23}_{k}=T^{13}_{j}-T^{24}_{k}
=e^{i\vec{\pi} \cdot \vec{j}}+e^{i\vec{\pi} \cdot \vec{k}}=0$
holds\cite{R5}.
The above 6 generators form
SO(4)$\sim$SU(2$)_S\otimes$SU(2$)_{\eta_Y}$\cite{YZ} whose generators are
$(\sigma^{12},\sigma^{34},\sigma^{14}-\sigma^{23},\sigma^{13}+\sigma^{24},
\eta^{x}_{Y},\eta^{y}_{Y})\sim(\vec{S},\vec{\eta}^Y)$
with
$\vec{\eta}_Y=(\eta^{x}_{Y},\eta^{y}_{Y},\eta^{z})$ as introduced in \cite{YZ}.

By a similar analysis (though it is clear from (\ref{2-3})) we find
the symmetry operators for $H^{S}_{Heisen.}$; global
$\sigma^{\mu\nu}\sim(\vec{S},\vec{\eta})$ and local $\gamma^{5}_{l}$.
The first 6 operators form O(4)$\sim$SU(2$)_S\otimes$SU(2$)_{\eta_Y}$
and, combined with $\gamma^5$ and 1, are enlarged to form
U(2$)_S\otimes$U(2$)_{\eta}$ whose generators are
$((\vec{S}, (1-\gamma^5)/2), (\vec{\eta}, (1+\gamma^5)/2))$.
But on the  $P_F$ restricted Hilbert space for this model this
reduces to the U(2$)_S$ symmetry.

For other models, the following commutation rules hold.
\begin{equation}
[\tilde{{\cal S}}, H_{EKS}]=
-2t\sum_{<jk>}\sum_{acd=1,2,3,4}
(x^{ac}_{j}-x^{cd}_{k})X^{ac}_{j}X^{cd}_{k}(-1)^{F(c)},
\end{equation}
\begin{equation}
[P\tilde{{\cal S}}P, \tilde{H}_{t-J}]=
-2t\sum_{<jk>}\sum_{acd=1,3,4}
(x^{ac}_{j}-x^{cd}_{k})X^{ac}_{j}X^{cd}_{k}(-1)^{F(c)}
=P[\tilde{{\cal S}}, H_{EKS}]P,
\end{equation}
\begin{equation}
P_F[\tilde{{\cal S}}, H^{S}_{Heisen.}]P_F=
-2t\sum_{<jk>}\sum_{acd=3,4}
(x^{ac}_{j}-x^{cd}_{k})X^{ac}_{j}X^{cd}_{k}(-1)^{F(c)}
=P_F[\tilde{{\cal S}}, H_{EKS}]P_F,
\end{equation}
\begin{equation}
P_B[\tilde{{\cal S}}, H^{\eta}_{Heisen.}]P_B=
-2t\sum_{<jk>}\sum_{acd=1,2}
(x^{ac}_{j}-x^{cd}_{k})X^{ac}_{j}X^{cd}_{k}(-1)^{F(c)}
=P_B[\tilde{{\cal S}}, H_{EKS}]P_B,
\end{equation}
Noting that
$X^{ac}$ with $a,c=1,2,3,4$,
$PX^{ac}P$ (or $X^{ac}$ with $a,c=1,3,4$),
$P_FX^{ac}P_F$ (or $X^{ff'}$ with $f,f'=3,4$),
and
$P_BX^{ac}P_B$ (or $X^{bb'}$ with $b,b'=1,2$)
form
U(2/2), U(1/2), U(2$)_{S}$ and U(2$)_{\eta}$
respectively where $X=\sum_{j=1}^{L}X_j$,
the symmetries of the models are clear.

\section*{ACKNOWLEDGMENT}

I would like to thank  Professor Bunji Sakita
 for  suggestions, discussions, and encouragement
throughout this work.
I also thank the Rotary Foundation for a scholarship.
This work was supported by NSF grant PHY90$-$20495 and
PSCBHE grant 6$-$63351.


\begin{thebibliography}{99}
\bibitem{R2} The 8 operators $X^{fb}_j$ and $X^{bf}_j$ with $b=1,2$
and $f=3,4$ are fermionic operators at each site.
\end{thebibliography}
\end{document}